\documentclass[pra,preprint,showpacs,superscriptaddress,showkeys]{revtex4-1}
\usepackage{amsmath,amssymb,amsthm,times,graphics,graphicx,bm}
\usepackage[colorlinks,linkcolor=red]{hyperref}

\newcommand{\be}{\begin{equation}}
\newcommand{\ee}{\end{equation}}
\newcommand{\ben}{\begin{eqnarray}}
\newcommand{\een}{\end{eqnarray}}

\def\ket#1{ | #1 \rangle}
\def\bra#1{{\langle #1 | }}
\def\tr{ {\rm{Tr }}}

\newcommand{\proj}[1]{\mbox{$|#1\rangle \!\langle #1 |$}}
\newcommand{\avg}[1]{\left\langle #1 \right\rangle}

\newtheorem{theorem}{Theorem}

\begin{document}
\title{On the negativity of random pure states}
\author{Animesh Datta}
 \email{a.datta@imperial.ac.uk}
 \affiliation{Institute for Mathematical Sciences, 53 Prince's Gate, Imperial College, London, SW7 2PG, UK}
 \affiliation{QOLS, The Blackett Laboratory, Imperial College London, Prince Consort Road, SW7 2BW, UK}

\date{\today}
\begin{abstract}
This paper deals with the entanglement, as quantified by the negativity, of pure quantum states chosen at random from the invariant Haar measure. We show that it is a constant ($0.72037$) multiple of the maximum possible entanglement. In line with the results based on the concentration of measure, we find evidence that the convergence to the final value is exponentially fast. We compare the analytically calculated mean and standard deviation with those calculated numerically for pure states generated via pseudorandom unitary matrices proposed by Emerson \emph{et. al.} [Science, \textbf{302}, 3098, (2003)]. Finally, we draw some novel conclusions about the geometry of quantum states based on our result.
\end{abstract}



\maketitle

\section{Introduction}

Entanglement has come to be believed as one of the cornerstones of quantum information science. The necessity of entanglement in quantum computation~\cite{jl03} and information~\cite{masanes06a,pw09} tasks are well acknowledged. Substantial amounts of experimental effort is expended in the generation and manipulation of quantum entanglement. Nevertheless, the role of entanglement in quantum information science in general, and quantum computation in particular, is far from clear. Meyer has presented a version of the quantum search algorithm that requires no entanglement~\cite{meyer00a}, and instances are known of mixed-state quantum computation where exponential speedup is attained in the presence of only limited amounts of entanglement~\cite{dfc05}, and other quantities have been proposed as alternate resources for the speedup~\cite{dsc08,dg09}. Recent results have further illuminated  the role of entanglement in pure-state quantum computation. It was already known, due to the Gottesman-Knill theorem~\cite{nielsen00a}, that entanglement is by no means sufficient for universal quantum computation. The new results~\cite{bmw09,gfe09} show that, in fact, almost all pure states are too entangled to be a universal resource for quantum computation. Though proved in the context to measurement-based quantum computation, and based on the geometric measure of entanglement, which is the absolute square of the inner product with the closest product state, these results drive home the point that implications on the lines of ``more entanglement implies more computational power" are fallacious~\cite{gfe09}. The strategy employed for proving these results can generally be termed as ``concentration of measure"~\cite{hlw06}, by which a typical pure state, chosen at random from the left- and right-invariant Haar measure, is almost always maximally entangled across any bipartition.

Arguments based on the concentration of measure have been used to obtain average value of measures of correlations and entanglement in typical quantum states. Concentration of measure is a very powerful concept from measure theory, which puts bounds on how much the values of certain smooth (Lipshitz) functions can vary from their mean value. This is a consequence of the remarkable fact that the uniform distribution of the $k$-sphere $\mathbb{S}^k$ is concentrated largely on the equator for large $k$, and any polar cap smaller than a hemisphere has a relative volume exponentially small in $k$. Examples in quantum information theory include the entropy of the reduced density matrix, entanglement of formation, distillable common randomness~\cite{hlw06}. The entropy of reduced density matrices of typical states has also been conjectured and calculated independently~\cite{page93,s96,fk94,ruiz95}, as has been their concurrence, purity and the linear entropy~\cite{scott03}. Not
much is however known of one of the most common and computable measures of entanglement, the negativity~\cite{zhsl98,vw02}, in random Haar distributed pure states. In this paper, our endeavor will be to address this question. We show that the negativity of a random pure state taken from a Haar distribution, is a constant multiple of the maximum possible. This entanglement can also be generated efficiently using two qubit gates~\cite{odp07}. We will evaluate this constant using techniques similar to those in Refs.~\cite{s96,scott03}, and confirm our results numerically using efficiently generated pseudorandom unitaries~\cite{emerson03a}. For simplicity, we will only present results for equal bipartitions, but the extensions to unequal splits is straightforward.

That the negativity (defined in Eq. (\ref{E:neg})) of random pure states is less than maximal might seem to contradict the statement that random pure states in large enough Hilbert spaces are close to being maximally entangled. This is, however, not true in general. As shown in~\cite{hlw06}, for a state residing in a Hilbert space of dimension $d_A \times d_B$ with a reduced state $\rho_A=\tr_B(\rho),$ and $d_B$ is a large enough multiple of $d_A\log d_A/\epsilon^2,$ then
 \be
 \label{E:cluster}
(1-\epsilon)\frac{1}{d_A}\mathbb{I} \leq \rho_A \leq
(1+\epsilon)\frac{1}{d_A}\mathbb{I}
 \ee
If a state satisfies Eq.~(\ref{E:cluster}), then its negativity is evidently near-maximal. But as the condition for its validity shows, this is only true when the bipartite split is quite asymmetrical. Thus, for equal bipartite splits, which is often of interest in quantum information science, there is no \emph{a priori} reason to expect the negativity of random pure states to be close to maximal. This is the case we study here. Just to highlight the degree of asymmetry needed to have the negativity close to maximal, for $\epsilon = 0.1$ and $d_A=2,$ we require
$d_B \gg 200,$ and for $d_A=16,$ $d_B \gg 6400.$

The outline of the paper is as follows. In Sec.~\ref{sec:mean}, we begin by deriving the expression of average negativity. It
involves performing integrations over the probability simplex which are rewritten in terms of other nonconstrained variables, finally leaving us with a combination of hypergeometric functions. Sec.~\ref{sec:Var} derives the expressions for the variance in the negativity in terms of similar hypergeometric functions. These functions are explicitly evaluated in Sec.~\ref{sec:evals} numerically. This is necessary as the series we have is provably not summable in closed form, which we discuss in brief in Appendix~\ref{app:digress}. We obtain the final expression for the average negativity of Haar-distributed random pure states. We also compare our results with a numerical simulation using pseudorandom
pure states generated from efficiently generated pseudorandom unitaries~\cite{emerson03a}, finding good agreement. We finally conclude in Sec.~\ref{sec:conclude} with discussions about the ramifications of our finding on the geometry of the set of quantum states. We also discuss the prospect of extending the present analysis to random mixed quantum states.

\section{Negativity of typical pure states}
\label{sec:mean}

The negativity is an entanglement monotone which is based on the partial transpose test of detecting entanglement~\cite{p96}. Given a bipartite quantum state residing in $\mathcal{H}_A\otimes \mathcal{H}_B$ with dimensions $\mu$ and $\nu$, called $\rho_{AB},$ the negativity is defined as
 \be
 \label{E:neg}
  \mathcal{N}(\rho_{AB}) = \frac{||\rho_{AB}^{T_A}||-1}{2},
 \ee
where $\rho_{AB}^{T_A}$ denotes the partial transpose with respect to subsystem $A$, and $||\sigma||$ denotes the trace norm, or sum of the absolute values of the eigenvalues of $\sigma,$ when $\sigma$ is Hermitian, as is the case with $\rho_{AB}^{T_A}.$ For pure states residing in the above space, it is always possible to write a Schmidt decomposition~\cite{nielsen00a}. This paper will only deal with the scenario $\mu=\nu$, the extension to the unequal case being tedious, but straightforward. The distribution of the Schmidt coefficients is given by (for $\mu = \nu$)~\cite{lp88}
    \be
P(\mathbf{p}) \mathrm{d}\mathbf{p} = N \delta(1-\sum_{i=1}^\mu
p_i) \prod_{1\leq i < j \leq \mu} (p_i-p_j)^2 \prod_{k=1}^\mu
\mathrm{d}\mathbf{p}_k,
    \ee
where $\delta(\cdot)$ is the Dirac delta function. The negativity for pure states is
    \be
\mathcal{N} =
\frac{1}{2}\left[\left(\sum_{i=1}^\mu\sqrt{p_i}\right)^2-1\right]
=\frac{1}{2}\mathop{\sum_{i,j=1}}_{i\neq j}^{\mu}\sqrt{p_i p_j}
\label{E:negdef}
    \ee
and its mean is given by
    \be
\avg{\mathcal{N}}= \frac{1}{2} \int \mathop{\sum_{i,j=1}}_{i\neq
j}^{\mu}\sqrt{p_i p_j}P(\mathbf{p}) \mathrm{d}\mathbf{p}.
    \ee
At the outset, it helps to change variables such that $q_i=r p_i$ which removes the hurdle of integrating over the probability simplex~\cite{s96,scott03}, whereby
    \be
Q(\mathbf{q})\mathrm{d}\mathbf{q}\equiv \prod_{1\leq i<j\leq
\mu}\left(q_i-q_j\right)^2
\prod_{k=1}^\mu e^{-q_k}\,\mathrm{d}q_k \label{Q}\\
=N\,e^{-r}r^{\mu^2-1}P(\mathbf{p})\,\mathrm{d}\mathbf{p}\,\mathrm{d}r\;.
    \ee
The new variables $q_i$ take on values independently in the range $[0,\infty),$ and $r$ is a scaling factor given by $r =\sum_iq_i$. Integrating over all the values of the new variables, we find that the normalization constant is given by $N=\overline Q/\Gamma(\mu\nu)$, where $\overline{Q}\equiv\int Q(\mathbf{q})d\mathbf{q}$. Similarly, we find that
\begin{equation}
\int \sqrt{q_i q_j}Q(\mathbf{q})\mathrm{d}\mathbf{q} = \overline Q\, \frac{\Gamma(\mu^2+1)}{\Gamma(\mu^2)} \int \sqrt{p_i p_j}P(\mathbf{p})\,\mathrm{d}\mathbf{p}\;, \label{QtoP}
\end{equation}
with $\Gamma(\mu)=(\mu-1)!.$ Notice that the first product in Eq.~(\ref{Q}) is the square of the Van der Monde determinant~\cite{s96,scott03}
 \be
\hspace{-2.0cm} \Delta(\mathbf{q}) \,\equiv\, \prod_{1\leq i<j\leq
\mu}\left(q_i-q_j\right) = \left|
\begin{array}{ccc}
 1 & \ldots & 1 \\
 q_1 & \ldots & q_\mu \\
 \vdots & \ddots & \vdots \\
 q_1^{\mu-1} & \ldots & q_\mu^{\mu-1}
 \end{array} \right|
 = \left| \begin{array}{ccc}
 L_0(q_1) & \ldots & L_0(q_\mu) \\
 L_1(q_1) & \ldots & L_1(q_\mu) \\
 \vdots & \ddots & \vdots \\
\Gamma(\mu) L_{\mu-1}(q_1)  & \ldots & \Gamma(\mu)L_{\mu-1}(q_\mu)
 \end{array} \right|\;.
 \label{Van2}
\ee
The second determinant in Eq.~(\ref{Van2}), follows from the basic property of
invariance after adding a multiple of one row to another, and the polynomials $L_k(q)$ judiciously chosen to be Laguerre
polynomials~\cite{gradshteyn}, satisfying the orthogonality
relation
\begin{equation}
\int_0^\infty dq\,e^{-q} L_k(q)L_l(q) = \delta_{kl}\;. \label{L1}
\end{equation}
These facts in hand, we can evaluate
\begin{eqnarray}
 \overline{Q} &=& \int\Delta(\mathbf{q})^2\prod_{k=1}^\mu e^{-q_k}\,dq_k \nonumber\\
              &=&\mathop{\sum_{T,R \in \mathcal{S}_{\mu}}} (-1)^{T+R}\prod_{k=1}^\mu\Gamma(T(k))\Gamma(R(k))\int dq_k\,e^{-q_k} L_{T(k)-1}(q_{k})L_{R(k)-1}(q_k) \nonumber \\
              &=&\sum_{R \in \mathcal{S}_{\mu}}(1)^R\prod_{k=1}^\mu\Gamma(R(k))^2=\mu!\prod_{k=1}^\mu\Gamma(k)^2\;,
\end{eqnarray}
with $T,R$ being elements of the permutation group on $\mu$
elements $\mathcal{S}_{\mu}.$ We can now calculate the integral over $\{q_1,\cdots,q_{\mu}\}$ in Eq. (\ref{QtoP}) as
 \ben
 &&\mathop{\sum_{i,j=1}}_{i\neq j}^\mu \int \sqrt{q_i q_j}Q(\mathbf{q})\,\mathrm{d}\mathbf{q} \nonumber\\
 &&=\mathop{\sum_{i,j=1}}_{i\neq j}^\mu \int \sqrt{q_i q_j}\prod_{m=1}^{\mu} dq_m e^{-q_m}\sum_{T,R \in \mathcal{S}_{\mu}}(-1)^{T+R}\prod_{m=1}^{\mu}\Gamma(T(m))\Gamma(R(m))L_{T(m)-1}(q_m)L_{R(m)-1}(q_m) \nonumber \\
 &&=\overline{Q}\mathop{\sum_{k,l=0}}^{\mu-1}\sum_{R \in \mathcal{S}_2}(-1)^{R}\int\sqrt{q_k q_l}L_{R(k)-1}(q_k)L_{k-1}(q_k)L_{R(l)-1}(q_l)L_{l-1}(q_l)e^{-q_k-q_l}dq_kdq_l \nonumber\\
 &&=\overline{Q}\sum_{k,l=0}^{\mu-1}\sum_{R \in \mathcal{S}_2} (-1)^{R}I_{k,R(k)}^{(1/2)}I_{l,R(l)}^{(1/2)}\nonumber\\
 &&= \overline{Q} \sum_{k,l=0}^{\mu-1}\left|\begin{array}{cc}
   I_{kk}^{(1/2)} & I_{kl}^{(1/2)} \\
   I_{lk}^{(1/2)} & I_{ll}^{(1/2)} \\
 \end{array} \right|,    \een
where
    \be
I_{kl}^{(\beta)} \equiv
\int_0^{\infty}e^{-q}q^{\beta}\,L_k(q)L_l(q)\;\mathrm{d}q,
\label{E:Int}
    \ee
$|\cdot|$ is the determinant and we have used the orthonormality condition in Eq. (\ref{L1}) in the first step of the evaluation. We thus have
    \be
 \label{E:avgneg}
\avg{\mathcal{N}}=\frac{1}{2\mu^2}\sum_{k,l=0}^{\mu-1}\left[
I_{kk}^{(1/2)}I_{ll}^{(1/2)}-\left(I_{kl}^{(1/2)}\right)^2\right]\;,
    \ee
except that the integral needs to be evaluated.

\section{Variance in the negativity}
\label{sec:Var}

Having calculated the mean of the negativity for random, Haar
distributed pure states, we move on to calculate its variance.
Based on the definition of negativity in Eq. (\ref{E:negdef}), we obtain the expression for the variance of the negativity as
($\sigma$ is the standard deviation)
    \be
\sigma^2 =
\frac{1}{4}\left[\avg{\left(\sum_{i=1}^\mu\sqrt{p_i}\right)^4} -
\avg{\left(\sum_{i=1}^\mu\sqrt{p_i}\right)^2}^2  \right].
    \ee
The second term has already been evaluated in the previous
section, so we need con concern ourselves with the first term. We begin by expanding the fourth power above as
 \be
 \hspace{-2.5cm}
\left(\sum_{i=1}^\mu\sqrt{p_i}\right)^4=1+
2\mathop{\sum_{i,j=1}}_{i\neq j}^{\mu} \sqrt{p_ip_j} +
2\mathop{\sum_{i,j=1}}_{i\neq j}^{\mu}p_ip_j +
4\mathop{\sum_{i,j,k=1}}_{i\neq j\neq k}^{\mu} p_i\sqrt{p_jp_k} +
\mathop{\sum_{i,j,k,l=1}}_{i\neq j\neq k \neq l}^{\mu}
\sqrt{p_ip_jp_kp_l}.
 \ee
Each of these terms can now be individually evaluated, and
omitting the details we just present the results as
 \ben
 \mathop{\sum_{i,j=1}}_{i\neq j}^{\mu}p_ip_j &=& \frac{1}{\mu^2(\mu^2+1)}\sum_{k,l=0}^{\mu-1}\left|\begin{array}{cc}
   I_{kk}^{(1)} & I_{kl}^{(1)} \\
   I_{lk}^{(1)} & I_{ll}^{(1)} \\
 \end{array} \right|,   \\
 \mathop{\sum_{i,j,k=1}}_{i\neq j\neq k}^{\mu} p_i\sqrt{p_jp_k} &=&  \frac{1}{\mu^2(\mu^2+1)}\sum_{k,l,m=0}^{\mu-1} \left|  \begin{array}{ccc}
   I_{kk}^{(1)} & I_{kl}^{(1/2)} & I_{km}^{(1/2)} \\
   I_{lk}^{(1)} & I_{ll}^{(1/2)} & I_{lm}^{(1/2)} \\
   I_{mk}^{(1)} & I_{ml}^{(1/2)} & I_{mm}^{(1/2)} \\
 \end{array}\right|, \\
 \mathop{\sum_{i,j,k,l=1}}_{i\neq j\neq k \neq l}^{\mu}\sqrt{p_ip_jp_kp_l}&=&
 \frac{1}{\mu^2(\mu^2+1)}\sum_{k,l,m,n=0}^{\mu-1} \left| \begin{array}{cccc}
   I_{kk}^{(1/2)} & I_{kl}^{(1/2)} & I_{km}^{(1/2)} & I_{kn}^{(1/2)} \\
   I_{lk}^{(1/2)} & I_{ll}^{(1/2)} & I_{lm}^{(1/2)} & I_{ln}^{(1/2)} \\
   I_{mk}^{(1/2)} & I_{ml}^{(1/2)} & I_{mm}^{(1/2)} & I_{mn}^{(1/2)} \\
   I_{nk}^{(1/2)} & I_{nl}^{(1/2)} & I_{nm}^{(1/2)} & I_{nn}^{(1/2)} \\
 \end{array} \right|\!.
 \een

\section{Evaluating the integrals}
\label{sec:evals}

Having derived formal expressions for the mean and standard
deviation of the negativity of a random pure state, we now need to evaluate the integral in Eq. (\ref{E:Int}). To that end, we use the generating function for Laguerre polynomials~\cite{gradshteyn}
    \be
 (1-z)^{-1} e^{xz/z-1} = \sum_{l=0}^{\infty}
 L_l(x)z^l\;\;\;\;\;\;\;\;  |z|\leq1,
    \ee
and
    \be
\hspace{-2.0cm}\int_0^{\infty}e^{-st}t^{\beta}\,L_n^{\alpha}(t)\;\mathrm{d}t=
\frac{\Gamma(\beta+1)\,\Gamma(\alpha+n+1)}{n!\,\Gamma(\alpha+1)}s^{-\beta-1}F\left(-n,\beta+1;\alpha+1,\frac{1}{s}\right),
    \ee
$F$ being the hypergeometric function such that
    \be
F(a,b;c;z)= \sum_{n=0}^{\infty}
\frac{(a)_n\,(b)_n}{(c)_n}\frac{z^n}{n!},
    \ee
and $(a)_n = a(a+1)(a+2)...(a+n-1)$ is the Pochhammer symbol. Note that if $a$ is a negative integer, $(a)_n = 0$ for $n > |a|$ and
the hypergeometric series terminates. Then,
    \ben
\sum_{l=0}^{\infty} I_{kl}^{(\beta)} z^l
&=&\int_0^{\infty}e^{-x}x^{\beta}\,L_k(x)(1-z)^{-1}
e^{xz/z-1}\;\mathrm{d}x  \nonumber\\
&=&
s\int_0^{\infty}e^{-sx}x^{\beta}\,L_k(x)\;\mathrm{d}x\;\;\;\;\;\;\;\;\;\;\;\;\;\;s=1/(1-z) \nonumber\\
&=&s^{-\beta}\Gamma(\beta+1)F\left(-k,\beta+1;1;\frac{1}{s}\right)\nonumber\\
&=&\Gamma(\beta+1)\sum_{t=0}^{k}\frac{(-k)_t\,(\beta+1)_t}{(1)_t}\frac{1}{t!}(1-z)^{t+\beta}\nonumber\\
&=&\Gamma(\beta+1)\sum_{l=0}^{\infty}\sum_{t=0}^{k}\frac{(-1)^l}{l!}\frac{(-k)_t\,(\beta+1)_t}{(t!)^2}(t+\beta)_{\underline{l}}\,z^l,
    \een
whereby
    \be
I_{kl}^{(\beta)}=\Gamma(\beta+1)\frac{(-1)^l}{l!}\sum_{t=0}^{k}\frac{(-k)_t\,(\beta+1)_t}{(t!)^2}(t+\beta)_{\underline{l}}\;,
    \ee
and $(a)_{\underline{n}}=a(a-1)(a-2)...(a-n+1)$ is the `falling factorial'. Using the following identities for the Pochhammer
symbols
    \ben
(x)_{\underline{n}}&=&(-1)^n (-x)_n,\\
(-x)_{n}&=& (-1)^n(x-n+1)_n, \\
(x)_n  &=& \Gamma(x+n)/\Gamma(x),
    \een
we have
   \ben
 \label{E:Ikl}
 I_{kl}^{(\beta)}&=&\frac{(-1)^l}{l!}\sum_{t=0}^{k}
\left(\begin{array}{c}
  k \\
  t \\
\end{array}
   \right)\frac{[\Gamma(t+\beta+1)]^2}{t!\,\Gamma(t-l+\beta+1)}\nonumber\\
&=&\frac{(-1)^l}{l!}\frac{\Gamma(1+\beta)^2}{\Gamma(1+\beta-l)} \;
_3F_2\left(\{\beta+1,\beta+1,-k\};\{1,\beta+1-l\};1\right).
 \een

        To get the final expression for the negativity in
Eq~(\ref{E:avgneg}), we substitute the expression for the
integrals from Eq~(\ref{E:Ikl}). The expressions are not very
illuminating, and for the lack of an asymptotic expression, we
present the numerical values in Table~(\ref{T:t1}), and plot them in Fig.~(\ref{negativitylimit}). See Appendix~\ref{app:digress} for a note on the summability of the series.  Anticipating a scaling in proportion to that of a maximally entangled state, we divide the mean expressed in Eq~(\ref{E:avgneg}) by the maximum possible negativity of a $\mu\times\mu$ system as $\mathcal{N}_{max}=(\mu-1)/2.$  As can be seen from Table~(\ref{T:t1}), the average value of the negativity saturates to a constant multiple of the maximum possible. This constant is found numerically, and in the asymptotic limit of large $n$, the negativity for an equal bipartition of a randomly chosen Haar-distributed pure state is
 \be
 \label{E:result}
\avg{\mathcal{N}} \sim 0.72037\left(\frac{2^{n/2}-1}{2}\right).
 \ee
Though we have not proven this analytically, it is easily seen
that the convergence is exponential. This can be concluded from the last column in the table, which shows the difference in the successive values of the third column. The value of $\Delta$ is progressively halved as the number of qubits $n$ goes up, and this shows that the negativity indeed saturates monotonically, and arguably, exponentially fast, to the value presented above. This is to be expected from the concentration of measure
results~\cite{hlw06}, which means that the negativity of random states in large enough Hilbert spaces is close to their
expectation value.

\begin{table}
\begin{center}
\begin{tabular}{c|c|r@{.}l|r@{.}l}
  \hline
  $n$ & $\mu$ & \multicolumn{2}{c|}{$\avg{\mathcal{N}}/\mathcal{N}_{max}$} & \multicolumn{2}{c}{$\Delta$}\\
  \hline
  \hline
   2  & 2    & 0&589049        \\
   4  & 4    & 0&65368  & 0&0646309 \\
   6  & 8    & 0&686614 & 0&0329346 \\
   8  & 16   & 0&703378 & 0&0167641 \\
  10  & 32   & 0&711878 & 0&0084994 \\
  12  & 64   & 0&716171 & 0&0042932 \\
  14  & 128  & 0&718332 & 0&0021611 \\
  16  & 256  & 0&719417 & 0&0010851 \\
  18  & 512  & 0&719961 & 0&0005439 \\
  20  & 1024 & 0&720233 & 0&0002724 \\
  22  & 2048 & 0&72037  & 0&0001366 \\
  \hline
\end{tabular}
\end{center}
\caption{Ratio of the negativity of random pure states to the
maximal negativity for Haar-distributed states of $n$ qubits. For an equipartition of $n$ qubit states, $\mu=2^{n/2}.$ $\Delta$ is the difference between successive values in the third column, providing evidence for an exponential convergence of $\avg{\mathcal{N}}/\mathcal{N}_{max}$ with $n$.} \label{T:t1}
\end{table}

\begin{figure}[!h]
\begin{center}
\resizebox{9.5cm}{6cm}{\includegraphics{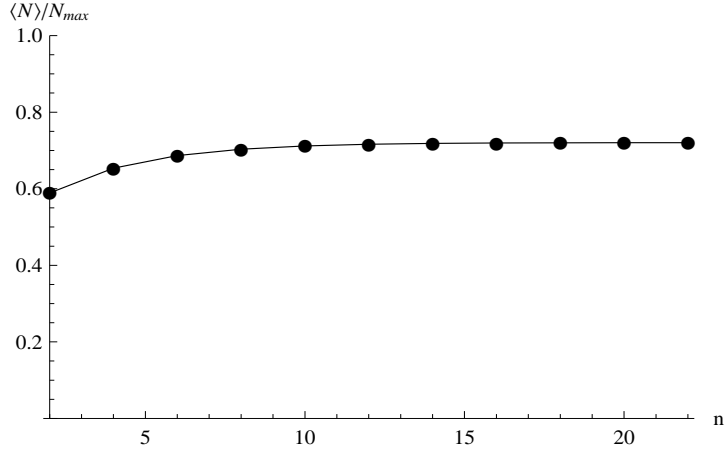}}
 \caption{The value of the normalized negativity
$\avg{\mathcal{N}}/\mathcal{N}_{max}$ of random Haar-distributed
pure states.}
 \label{negativitylimit}
\end{center}
\end{figure}

\subsection{Numerical verification}

\begin{figure}[!htb]
\begin{center}
\resizebox{16.5cm}{5cm}{\includegraphics{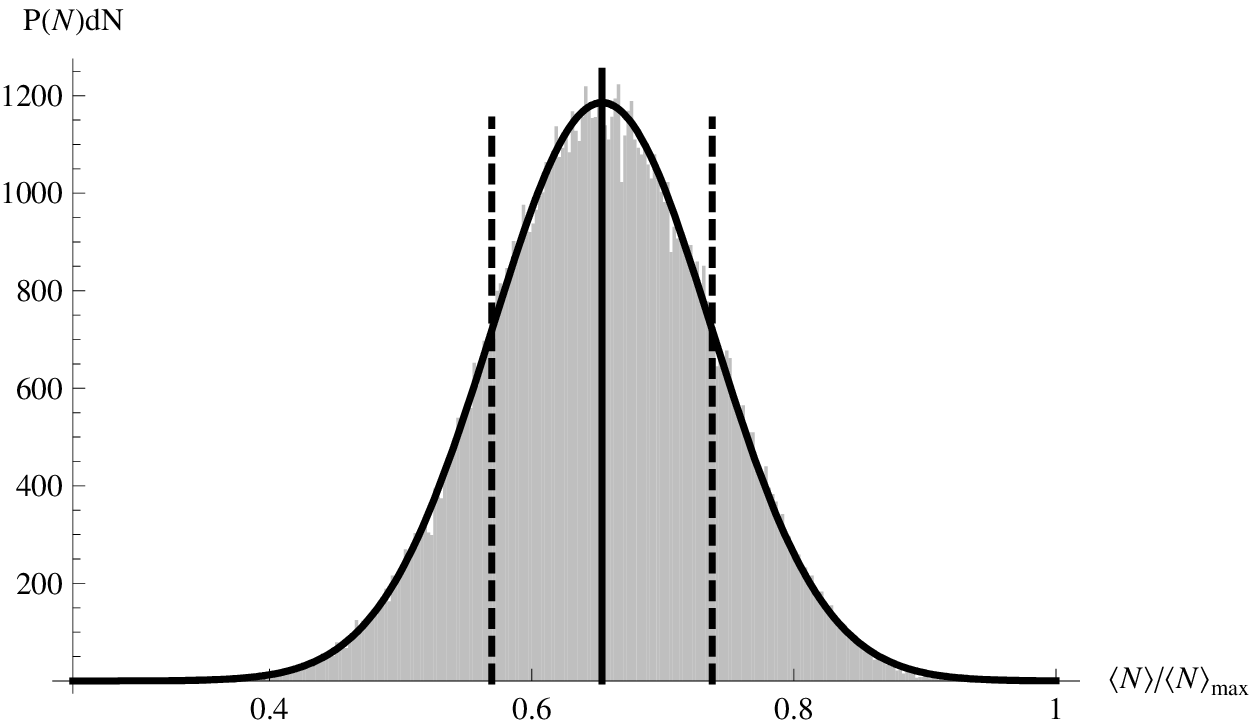}\includegraphics{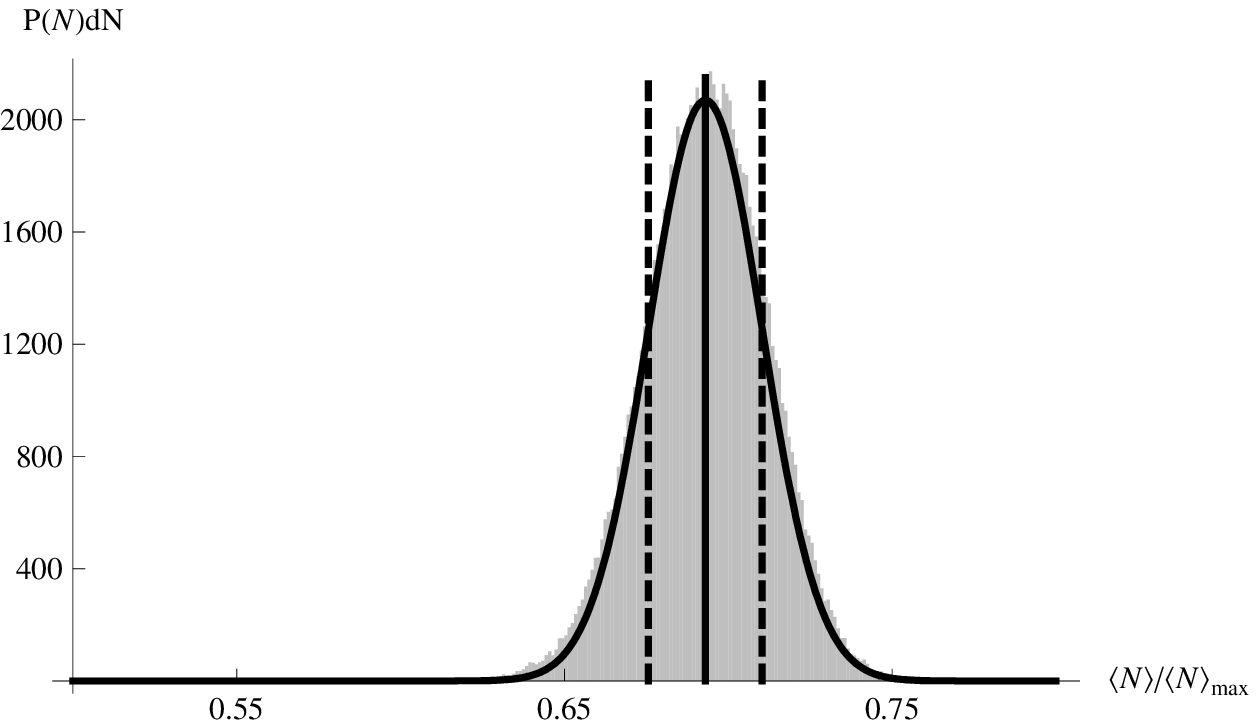}}
 \caption{Distribution of the negativity of $100000$ pseudorandom pure states, with $n=4$ (Left) and $n=8$ (Right). The pseudorandom unitaries used were generated via the techniques of \cite{emerson03a}, with $j=40$ interactions applied for each unitary. Also plotted is the gaussian distribution function with just the first two moments, as given by Eq.~(\ref{E:distfn}), as well as the analytically calculated mean (solid vertical line) and the standard deviations (dashed vertical lines). Although the convergence of the pseudorandom construction of~\cite{emerson03a} to the Haar measure is not obvious, it has been shown to do so~\cite{ell05,dop07}.}
 \label{F:numerics}
\end{center}
\end{figure}

As a final corroboration of our results, we test our calculations against numerically generated pure states. These are pseudorandom rather than random Haar-distributed. They are generated by applying pseudorandom unitaries presented in Ref.~\cite{emerson03a} on fiducial pure states. The negativity of these pure states is calculated and plotted as a histogram in Fig.~(\ref{F:numerics}). We compare this to an approximation of the cumulant generating function, and the probability distribution function for the negativity itself $P(\mathcal{N})d\mathcal{N}$, given by
 \ben
P(\mathcal{N})d\mathcal{N}&=&\frac{1}{2\pi}\int_{-\infty}^{\infty}d\omega
\exp\left(-i\mathcal{N}\omega +
\frac{\avg{\mathcal{N}}}{\mathcal{N}_{max}}i\omega +
\frac{\sigma}{\mathcal{N}_{max}}\frac{(i\omega)^2}{2!}\right)d\mathcal{N} \nonumber\\
&=&
\frac{1}{\sqrt{2\pi\sigma'^2}}e^{-\left(\mathcal{N}-\mathcal{N'}\right)^2/2\sigma'^2}d\mathcal{N},
\label{E:distfn}
 \een
where $\mathcal{N'}=\avg{\mathcal{N}}/\mathcal{N}_{max}$ and
$\sigma'=\sigma/\mathcal{N}_{max}.$ As is evident from
Fig.~(\ref{F:numerics}), the distributions are very localized, and the gaussian distribution seems quite apt.

\section{Concluding Discussions}
\label{sec:conclude}

The negativity provides upper bounds on the teleportation capacity of a state, and its distillability, the latter via the logarithmic negativity. It is in these two contexts that our results on the negativity provides new insights, not achieved through other measures. To address the teleportation capacity, the singlet distance was introduced in~\cite{vw02}. It is defined as closest distance any quantum state can get to the singlet (the ideal resource for teleportation) while undergoing only local operations. Mathematically,
    \be
    \Delta(\ket{\Phi},\rho) = \inf_P||\proj{\Phi}-P(\rho)||_1
    \ee
where $P$ is the set of all local protocols, and $\ket{\Phi}$ is the singlet residing in $\mathbb{C}^m\otimes \mathbb{C}^m$. Note that in our case $m=2^{n/2} = 2\mathcal{N}_{max}+1.$ The following result, also proved in~\cite{vw02}
    \be
 \Delta(\ket{\Phi},\rho) \geq 2\left(1-\frac{2\mathcal{N}(\rho)+1}{m}\right)
    \ee
then immediately leads to the conclusion that a pure quantum state $\ket{\psi}$, chosen at random from the Haar measure, will with high probability have a singlet distance given (all $\approx$ signs here and henceforth apply to large $n$)
 \be
 \label{E:singletbnd}
\Delta(\ket{\Phi},\ket{\psi}) \geq
2\left(1-\frac{2\avg{\mathcal{N}}+1}{2^{n/2}}\right) \approx
2\left(1-\frac{\avg{\mathcal{N}}}{\mathcal{N}_{max}}\right)
\approx 0.55926,
 \ee
where we have used Eq. (\ref{E:result}), which is that $\frac{\avg{\mathcal{N}}}{\mathcal{N}_{max}} = 0.72037  =c < 1.$ This gives us a nontrivial lower bound on how close a typical pure state can be taken to a singlet by purely local operations. This can be recast in terms of an upper bound on the teleportation fidelity~\cite{hhh99,vw02} of random pure states as
  \be
  \label{E:fid}
 f_{opt} \equiv \max_{P}\bra{\Phi}P(\proj{\psi})\ket{\Phi} \leq \frac{2\avg{\mathcal{N}}+1}{m} \lesssim \frac{\avg{\mathcal{N}}}{\mathcal{N}_{max}} \approx 0.72037.
  \ee

Another application of our result can be found by using the logarithmic negativity~\cite{p05} as an upper bound on the entanglement of distillation $E_D(\rho)$. It was shown~\cite{vw02} that
 \be
 E_D(\rho) \leq E_{\mathcal{N}}(\rho)
 \ee
where $E_\mathcal{N}(\rho) = \log_2||\rho^{T_A}||_1 =
\log_2(2\mathcal{N}(\rho)+1).$ Using this, we get (where c =
0.72037, as after Eq. (\ref{E:singletbnd}))
 \be
\avg{||\rho^{T_A}||_1} = c\;2^{n/2} + 1-c,
 \ee
whereby for a pure state $\ket{\psi}$ chosen at random from the Haar measure, we can set the upper bound of distillable
entanglement to be
    \be
    \label{E:distent}
 E_D(\ket{\psi}) \leq \log_2\left(\avg{||\rho^{T_A}||_1}\right) \approx \frac{n}{2} +\log_2c.
    \ee
For the constant we present in this work, this provides us with a bound that is tighter by about half an ebit ($\log_2c \approx -0.47319$). Also note that we have taken a logarithm of the average, which is always greater than or equal to the average of the logarithm.

In addition to the obvious conclusions that the fidelity of teleportation and distillability of random pure states have nontrivial upper bounds, the above two mathematical results tell us a few things about the structure of the set of pure quantum states in general. Firstly, although a random pure state is very likely to be highly entangled (close to maximal), it is in no way close to the singlet state, at least in trace norm. This means that a nonzero fraction of these ``close to maximally entangled" states contain inequivalent types of entanglement which are not related by SLOCC operations to the canonical maximally entangled (singlet) state. A second, and probably stronger statement is that not only do random pure states lie in different inequivalent sets of maximally entangled states, but also that some of these classes have a greater ability to retain their entanglement under distillation protocols than others, resulting thereby in an overall lower distillation rate.

This paper shows that the negativity of $n$-qubit random pure states chosen from the Haar measure is a constant multiple of maximum possible negativity, which goes as $2^{n/2}$ for an equal bipartition of the state. We also provide evidence that the convergence to the asymptotic value is monotonic and exponentially fast. The value of the constant was not evaluated in closed form, and we showed why this was the case. The expression for the negativity is a sum of hypergeometric terms, and the techniques of creative telescoping show that our particular series in not summable. Finally, we show that the results of our analytic calculation are borne out by random states generated by applying pseudorandom unitaries on
fiducial states. We also show that probability distribution for the negativity is well approximated by a gaussian distribution
whose mean and variance we obtain analytically.

One issue that we have not addressed here is the extension of the above calculation to random quantum states that are mixed. This is made somewhat challenging by the fact that there does \emph{not} exist a unique measure on the space of mixed quantum states. Since any pure state can be generated by applying a unitary matrix on a fiducial state, a unique measure on the space of pure states can be derived from that on the space of unitary matrices, which is the rotationally invariant Haar measure. Mixed quantum states cannot be generated in a likewise manner, and therefore, it is not possible to capture the distribution of mixed states via the Haar measure. However, any mixed state can be diagonalized by a unitary matrix, and this motivates a product measures on the space of mixed states $\mathcal{M},$ which can be defined as $\mathcal{M}=\mathcal{E}\times P,$ where $P$ is the
usual Haar measure that captures the distribution of eigenvectors of the states. $\mathcal{E}$ is meant to capture the distribution of eigenvalues, and there is no unique way of doing that. Attempts have been made~\cite{zs01}, and the mean entanglement, as quantified, for instance, by the purity has been calculated, as has been the logarithmic negativity for tripartite states using minimal purifications~\cite{dop07}. The calculation of the negativity for states of this form will be the subject of a future publication. This will provide us with information about the typical entanglement(negativity) content of random mixed states, which are more and more likely to be encountered as we move closer to realistic implementations of quantum technology.

\appendix

\section{A mathematical digression}
\label{app:digress}

The final expression for the negativity, though seemingly compact, is, in fact a sum of exponentially many terms. This retards the evaluation of the quantities in Table~(\ref{T:t1}) drastically, unless a closed form is found for quantity in Eq. (\ref{E:Ikl}). Consequently, it would not only be
interesting, but indeed essential to have a closed form of the above expression. For some special instances of $k,l$ and $\beta$, this is possible. Unfortunately, this is not possible for general values of $k$ and $l$ (this paper deals only with $\beta = 1/2, 1$). In fact, it can be shown that there exists \emph{no} closed form solution for the sum in Eq. (\ref{E:Ikl}). The arguments leading to this `tragic' conclusion are presented next.

\begin{theorem}[Zeilberger's algorithm or the method of creative telescoping~\cite{pwz97}] Let $F(n,k)$ be a proper hypergeometric term. Then F satisfies a nontrivial recurrence of the form
$$
\sum_{j=0}^J a_j(n)F(n+j,k)=G(n,k+1)-G(n,k),
$$ in which $G(n,k)/F(n,k)$ is a rational function of n and k.
\end{theorem}
That this theorem applies to the sum we have at hand is evident. The application of this algorithm to the expression in Eq.~(\ref{E:Ikl}) yields third order recurrences which can be solved using the Gosper-Petkov\v{s}ek algorithm~\cite{pwz97,petkovsek}. This algorithm (also called
\texttt{Hyper}~\cite{footnote}) provides a complete solution to the problem in the sense that it either provides all the solution to the recurrence problem. On the other hand, the failure of the algorithm to come up with a solutions proves that the initial series \emph{cannot} be summed
into a closed form. It is the latter that happens in our case, thereby proving that the series in Eq.~(\ref{E:Ikl}) is not summable in closed form.

\section*{Acknowledgments}

It is a pleasure to thank Colston Chandler, Anil Shaji, Adolfo del Campo and Miguel Navascu\'{e}s for several interesting discussions during the course of this work, and Martin B. Plenio for several comments on the manuscript. AD was supported by EPSRC (Grant No. EP/C546237/1), EPSRC QIP-IRC and the EU Integrated Project (QAP).


\begin{thebibliography}{99}
\bibitem{jl03} R. Jozsa, N. Linden, Proc. Roy. Soc. A \textbf{459}, 2011, (2003).
\bibitem{masanes06a} L. Masanes, Phys. Rev. Lett. \textbf{96}, 150501, (2006).
\bibitem{pw09} M. Piani, J. Watrous, Phys. Rev. Lett. \textbf{102}, 250501 (2009).
\bibitem{meyer00a} D. A. Meyer, Phys. Rev. Lett. \textbf{85}, 2014, (2000).
\bibitem{dfc05} A. Datta, S. T. Flammia, C. M. Caves, Phys. Rev. A \textbf{72}, 042316, (2005).
\bibitem{dsc08} A. Datta, A. Shaji, C. M. Caves, Phys. Rev. Lett. \textbf{100}, 050502, (2008).
\bibitem{dg09} A. Datta, S. Gharibian, Phys. Rev. A \textbf{79}, 042325, (2009).
\bibitem{nielsen00a} M. A. Nielsen, I. L. Chuang, \textit{Quantum Computation and Quantum Information}, Cambridge Univ. Press, (2000).
\bibitem{bmw09} M. J. Bremner, C. Mora, A. Winter, Phys. Rev. Lett. \textbf{102}, 190502 (2009).
\bibitem{gfe09} D. Gross, S. T. Flammia, J. Eisert, Phys. Rev. Lett. \textbf{102}, 190501 (2009) .
\bibitem{hlw06} P. Hayden, D.W. Leung, A.Winter, Commun. Math. Phys. \textbf{265}, 95, (2006).
\bibitem{page93} D. Page, Phys. Rev. Lett. \textbf{71}, 1291, (1993).
\bibitem{s96} S. Sen, Phys. Rev. Lett. \textbf{77}, 1, (1996).
\bibitem{fk94} S. K. Foong, S. Kanno, Phys. Rev. Lett. \textbf{72}, 1148, (1994).
\bibitem{ruiz95} J. Sanchez-Ruiz, Phys. Rev. E \textbf{52}, 5653, (1995).
\bibitem{scott03} A. J. Scott and C. M. Caves, J. Phys. A \textbf{36}, 9553, (2003).
\bibitem{zhsl98} K. \.{Z}yczkowski, P. Horodecki, A. Sanpera, and M. Lewenstein, Phys. Rev. A \textbf{58}, 883, (1998).
\bibitem{vw02} G. Vidal, R. Werner, Phys. Rev. A \textbf{65}, 032314, (2002).
\bibitem{odp07} R.Oliveira, O.C.O.Dahlsten and M.B.Plenio, Phys. Rev. Lett \textbf{98}, 130502 (2007).
\bibitem{emerson03a} J. Emerson, Y. S. Weinstein, M. Saraceno, S. Lloyd and D. G. Cory, Science, \textbf{302}, 3098, (2003).
\bibitem{dop07} O.C.O. Dahlsten, R. Oliveira and M.B. Plenio J. Phys. A \textbf{40}, 8081, (2007).
\bibitem{p96} A. Peres, Phys. Rev. Lett. \textbf{77}, 1414, (1996).
\bibitem{lp88} S. Lloyd, H. Pagels, Ann. Phys. \textbf{188}, 186, (1988).
\bibitem{gradshteyn}I. S. Gradshteyn, I. M. Ryzhik, \textit{Table of Integrals, Series and Products}, Academic, New York, (1980).
\bibitem{hhh99} M. Horodecki, P. Horodecki, and R. Horodecki, Phys. Rev. A \textbf{60}, 1888, (1999).
\bibitem{ell05} J. Emerson, E. Livine and Seth Lloyd, Phys. Rev. A \textbf{72}, 060302(R), (2005).
\bibitem{vollbrechtwolf02}  K. G. H. Vollbrecht, and M. M. Wolf, J. Math. Phys. \textbf{43}, 4299, (2002).
\bibitem{p05} M. B. Plenio, Phys. Rev. Lett \textbf{95} 090503, (2005).
\bibitem{pwz97}M. Petkov\v{s}ek, H. S. Wilf, and D. Zeilberger, \textit{A=B}, A K Peters Ltd, (1997). Available for free download at
 \url{http://www.math.upenn.edu/~wilf/AeqB.html}.
\bibitem{petkovsek} M. Petkov\v{s}ek, J. Symb. Comp. \textbf{11}, 1, (1998).
\bibitem{zs01} K. \.{Z}yczkowski and H-J. Sommers, J. Phys. A \textbf{34}, 7111, (2001).
\bibitem{footnote} The codes for executing this and Zielberger's algorithm are available at \url{http://www.risc.uni-linz.ac.at/research/combinat/software/}.

\end{thebibliography}
\end{document}